\documentclass[lettersize,journal]{IEEEtran}
\usepackage{amsmath,amsfonts}
\usepackage{algorithmic}
\usepackage{algorithm}
\usepackage{array}
\usepackage{textcomp}
\usepackage{stfloats}
\usepackage{url}
\usepackage{verbatim}
\usepackage{graphicx}
\usepackage{cite}
\usepackage{subcaption}
\usepackage{caption}
\usepackage{overpic}
\usepackage{xcolor}
\begin{document}

\title{Low-Frequency Stable Single-Source Integral Equation}

\author{Alessandro Zuccotti, \IEEEmembership{Student Member,~IEEE}, Cedric M{\"u}nger, \IEEEmembership{Member,~IEEE}, Chien V. Le, ~\IEEEmembership{Member,~IEEE}, Paul~Olyslager, \IEEEmembership{Student Member,~IEEE}, and Kristof Cools, \IEEEmembership{Member, IEEE}
\thanks{This work was supported by European Research Council (ERC) under European Union’s Horizon 2020 Research and Innovation Program under Grant 101001847 and by Special Research Fund (BOF) of Ghent University under Grant BOF.PDO.2024.0016.01 .}
\thanks{The authors are with IDLab, Department of Information Technology at Ghent University - imec, Ghent, Belgium (Corresponding author: Alessandro Zuccotti e-mail: alessandro.zuccotti@ugent.be). This work has been submitted to the IEEE for possible publication. Copyright may be transferred without notice, after which this version may no longer be accessible.}}



\maketitle

\begin{abstract}
Among the integral equations for scattering by homogeneous dielectric objects, single-source integral equations (SSIEs) provide the advantage of having only a single electric or magnetic current unknown. A well-conditioned SSIE can be obtained by appropriately combining the interior and exterior Calder\'on formulas. The condition number of the resulting operator appearing in the well-conditioned SSIE remains bounded at low frequencies. However, SSIEs still suffer from low-frequency cancellation, causing loss of significant digits in the solution, the right-hand side, and the system matrix. In this work, we apply a rescaling strategy based on quasi-Helmholtz projectors to the well-conditioned SSIE. Primal and dual space projectors are required since both primal and dual spaces are used to discretize the SSIE operators. We show that the solution of the rescaled SSIE is immune to catastrophic cancellations at low frequencies. 
Numerical experiments corroborate the method by testing the accuracy of the radiated field for extremely low frequencies.
\end{abstract}

\vspace{-0.3cm}
\section{Introduction}
\IEEEPARstart{S}{cattering} of time-harmonic electromagnetic fields by homogeneous dielectric media can be efficiently modeled via boundary integral equations (BIEs). 
Dual-source formulations, such as the Poggio-Miller-Chang-Harrington-Wu-Tsai (PMCHWT) equation \cite{poggio1970integral, Chang_Harrington_PMCHWT, Wu_Tsai_PMCHWT} and the M\"uller equation \cite{muller2013foundations}, have been widely studied. 
On the other hand, single-source integral equations (SSIEs) \cite{Marx1984, Glisson1984,costabel2011kleinman} provide an attractive alternative with only a single equivalent auxiliary electric or magnetic current unknown.   
This auxiliary unknown is not a physical trace; the traces are recovered by applying the interior Calderón operators to it.
By appropriately combining the interior and exterior Calder\'on formulas, a well-conditioned SSIE can be obtained \cite{Valdes2011}. SSIE formulations have also been proposed to remedy the high-contrast breakdown \cite{GHV+2018,GVR2019}.
SSIEs involve products of Maxwell boundary integral operators, which require primal and dual basis functions for accurate discretization \cite{Munger_SSIE_2024_penetrable_struct}. Rao-Wilton-Glisson (RWG)\cite{rao_electromagnetic_1982} and Buffa-Christiansen (BC)\cite{BUFFA2005461} basis functions are typically chosen as the primal and dual spaces, respectively.
Although SSIEs can be made well-conditioned at low frequencies, they still suffer from numerical cancellation at very low frequencies, as is common in electromagnetic BIEs, due to the different scaling of the loop and star components of the unknown. Unlike conventional BIEs, in the SSIE formulation, this issue affects not only secondary quantities, such as the far field, but also the computation of the physical electric and magnetic traces recovered from the auxiliary unknown. 

This work presents a rescaled SSIE that yields the correct solution upon applying a number of Krylov iterations that is nearly independent of the frequency in the low-frequency regime. The loop and star components are rescaled in both primal and dual spaces using quasi-Helmholtz projectors.
These have been used to rescale the electric field integral equation (EFIE) \cite{ACB+2013,BCA2015,BCA2015b,HEA+2023}, the magnetic and combined field integral equations (MFIE and CFIEs) \cite{Merlini2020_lf_CFIE,LCA+2024}, the PMCHWT equation \cite{Beghein2017_lf_PMCHWT,GAM+2017,GSM+2025,LMA+2026}, and the M\"uller equation \cite{LC2026}.
We extend the rescaling scheme to a conforming SSIE discretization. As in the rescaled PMCHWT \cite{Beghein2017_lf_PMCHWT}, and unlike in the EFIE and the CFIE, the structure of the SSIE requires each discretized integral operator to be rescaled in its expansion and range spaces, i.e., the space in which the image of the operator is expanded after applying the inverse Gram matrix. This ensures the correct scaling of both the auxiliary and the physical unknowns, while preserving the favorable condition number in the low-frequency regime.
As a consequence of this rescaling, the rescaled SSIE retains the static null-space that characterizes the well-conditioned SSIE and the MFIE on multiply connected geometries. Numerical errors might therefore appear in the auxiliary solution, although we show that this does not propagate to the physical traces and the far field.
Unlike in the works cited above, on multiply connected geometries, SSIE rescaling requires separate global- and local-loop projectors in the projected double-layer operator. This preserves an identity of the static double-layer operator that would otherwise be satisfied only up to quadrature precision, leading to significant errors in low-frequency simulations.
\vspace{-0.25cm}
\section{Notation and Background}
Let $\Omega_1 \subset \mathbb{R}^3 $ be a connected domain filled with a homogeneous medium with permittivity $\epsilon_1=\epsilon_r\epsilon_0$ and permeability $\mu_1=\mu_r\mu_0$, and $\Omega_0:=\mathbb{R}^3\setminus\overline{\Omega_1} $ be the exterior domain, with permittivity $\epsilon_0$ and permeability $\mu_0$. We consider a time-harmonic incident electromagnetic field $(\mathbf{e}^{\mathrm{inc}}(\mathbf{r}),\mathbf{h}^{\mathrm{inc}}(\mathbf{r}))$ at frequency $\omega$ (with time dependence $e^{j\omega t}$). The Maxwell single- and double-layer integral operators on the boundary $\Gamma:=\partial\Omega_1$ are
\begin{equation}
    \begin{split}
        \mathcal{T}_{\kappa_i}&=\mathcal{T}_{s,{\kappa_i}}+\mathcal{T}_{h,{\kappa_i}},\\
            (\mathcal{T}_{s,{\kappa_i}}\ \mathbf{j})(\mathbf{r})&=-
        j\kappa_i\mathbf{\hat{n}}\times\int_\Gamma\frac{e^{-j\kappa_i R}}{4\pi R}\mathbf{j}(\mathbf{r'})dS',\\(\mathcal{T}_{h,{\kappa_i}}\ \mathbf{j})(\mathbf{r})&=\frac{1}{j\kappa_i}\mathbf{\hat{n}}\times\mbox{p.v.}\int_\Gamma\nabla\frac{e^{-j\kappa_i R}}{4\pi R}\nabla' \cdot\mathbf{j}(\mathbf{r'})dS',\\
            (\mathcal{K}_{\kappa_i}\ \mathbf{j})(\mathbf{r})&=\mathbf{\hat{n}}\times \mbox{p.v.}\int_\Gamma\nabla  \frac{e^{-j\kappa_i R}}{4\pi R}\times \mathbf{j}(\mathbf{r'})dS',
    \end{split}
\end{equation}
with $R=|\mathbf{r}-\mathbf{r'}|$, $\mathbf{\hat{n}}$ the outward surface normal on $\Gamma$, and $\kappa_i=\omega \sqrt{\epsilon_i\mu_i}$ the wavenumber in the domain $\Omega_i$.
The characteristic impedance is $\eta_i=\sqrt{\frac{\mu_i}{\epsilon_i}}$.
Assuming only external sources, by multiplicatively combining the interior and exterior Calderón formulas and exploiting their redundancy, the well-conditioned SSIE is obtained as
\begin{equation}\label{singlesource}
    \begin{split}
           \bigg[\Big(\mathcal{K}_{\kappa_0}\!-\frac{1}{2}\Big)\Big(-\mathcal{K}_{\kappa_1}\!+\frac{1}{2}\Big) \!+\frac{\eta_0}{\eta_1}\mathcal{T}_{\kappa_0}\mathcal{T}_{\kappa_1}\bigg]\ \!  \mathbf{p}\!=\mathbf{\hat{n}} \!\times \mathbf{e}^{\mathrm{inc}}.
    \end{split}
\end{equation}
A detailed derivation of \eqref{singlesource} can be found in \cite{Valdes2011} and \cite{Zuccotti2024}.
The unknown $\mathbf p$ in \eqref{singlesource} does not represent a physical trace. 
Applying the interior operators yields the electric and magnetic traces
\begin{equation}
\mathbf{\hat{n}} \times \mathbf{h}=-\frac{1}{\eta_1}\mathcal{T}_{\kappa_1} \mathbf{p},\quad \mathbf{e}\times\mathbf{\hat{n}} =\Big(-\mathcal{K}_{\kappa_1}+\frac{1}{2}\Big)\mathbf{p}.
\end{equation}

\section{Discretization}
Let $\mathbf{f}$ and $\mathbf{g}$ denote the RWG and BC bases, respectively.
The auxiliary unknown $\mathbf p$ in \eqref{singlesource} is expanded in the RWG space
\begin{equation}\label{RWGexp}
\mathbf p(\mathbf r)= \sum_i (\mathbf p^f)_i\mathbf{f}_i(\mathbf r),
\end{equation}
where $\mathbf p^f$ is the unknown vector of coefficients.
Each operator in \eqref{singlesource} is discretized using RWG or BC trial functions and rotated RWG or BC test functions. We denote by
$(\mathcal{O}_{xy})_{mn}=\langle \mathbf{\hat{n}} \times \mathbf{x}_m, \mathcal{O}\mathbf{y}_n\rangle$,
the matrix associated with the operator $\mathcal O$, where $\langle\cdot,\cdot\rangle$ is the $L^2(\Gamma)$ inner product. We also define the mixed Gram matrix $(N_{xy})_{mn}=\langle \mathbf{\hat{n}} \times \mathbf{x}_m, \mathbf{y}_n\rangle$. We denote its inverse by $N^{yx}$. The anti-symmetric property $N^{xy}=-{N^{yx}}^T$ applies. Equation \eqref{singlesource} is discretized as follows:

\begin{equation}\label{discretized_singlesource}
\begin{split}
&N^{fg}\bigg[\Big(\mathcal{K}_{\kappa_0}-\frac{1}{2}\Big)_{gf}N^{fg} \Big(-\mathcal{K}_{\kappa_1}+\frac{1}{2}\Big)_{gf}+\\
&+\frac{\eta_0}{\eta_1}(\mathcal{T}_{\kappa_0})_{gg}({N^{gf}})(\mathcal{T}_{\kappa_1})_{ff}\bigg]\ \mathbf{p}^f
=N^{fg}(\mathbf{\hat{n}} \times \mathbf{e}^{\mathrm{inc}})_g,
\end{split}
\end{equation}
with $((\mathbf{\hat{n}} \times \mathbf{e}^{\mathrm{inc}})_g)_i=\langle \mathbf{\hat{n}} \times \mathbf{g}_i,\mathbf{\hat{n}} \times \mathbf{e}^{\mathrm{inc}}\rangle$, and $\Big(\mathcal{K}_{\kappa_i}-\frac{1}{2}\Big)_{gf}=\Big((\mathcal{K}_{\kappa_i})_{gf}-\frac{1}{2}N_{gf}\Big)$.
This conforming scheme yields an accurate discretization of all the integral operators \cite{CoolsMFIEconforming}, and preserves the spectral properties of the system.   
The physical traces are recovered as 
\begin{equation}
\label{physical_traces_discretized}
\begin{split}
(\mathbf{\hat{n}} \times \mathbf{h})^g&=-\frac{1}{\eta_1}{N^{gf}}(\mathcal{T}_{\kappa_1})_{ff}\mathbf{p}^f,\\
(\mathbf{e}\times\mathbf{\hat{n}})^f&=N^{fg}\big(-\mathcal{K}_{\kappa_1}+\tfrac{1}{2}\big)_{gf}\mathbf{p}^f.
    \end{split}
\end{equation}

\section{Low-frequency cancellation}
At low frequencies, numerical instabilities arise in the evaluation of \eqref{discretized_singlesource}. 
To analyze the frequency behavior of the involved quantities, it is convenient to consider their loop-star decomposition. Coefficient vectors in the RWG and BC spaces admit the quasi-Helmholtz decomposition \cite{vecchiLoopStar,Beghein2017_lf_PMCHWT}
\begin{equation*}
    \begin{split}
             \mathbf j^f= \Lambda_f \mathbf l_f+ H_f\mathbf h_f+\Sigma_f\mathbf s_f,\quad
            \mathbf j^g= \Lambda_g \mathbf l_g+ H_g\mathbf h_g+\Sigma_g\mathbf s_g.
    \end{split}
\end{equation*}
The matrices $(\Lambda_x,H_x,\Sigma_x)$ span the local loop (solenoidal), the global loop (harmonic), and the star subspaces, respectively \cite{Beghein2017_lf_PMCHWT,Cools_nullsp_MFIE}. For the proposed rescaling strategy, we will need only the definitions of $\Sigma_f$ and $\Lambda_f$ from \cite{Beghein2017_lf_PMCHWT,Merlini2020_lf_CFIE},
with one column removed to account for the corresponding rank deficiency. 
These are the signed connectivity matrices between edges and vertices, and between edges and cells. We define
\begin{equation}
\begin{split}
\!(\Lambda_f)_{mn} &=
\begin{cases}
+1, & \text{if}\ v_n=v_m^+,\\
-1, & \text{if}\ v_n=v_m^-,\\
0, & \text{otherwise},
\end{cases}
\ \begin{array}{l}
m=1,\ldots,N_{\mathrm{e}},\\
n=1,\ldots,N_{\mathrm{v}}-1,
\end{array}\\
(\Sigma_f)_{mn} &=
\begin{cases}
+1, & \text{if}\ c_n=c_m^+,\\
-1, & \text{if}\ c_n=c_m^-,\\
0, & \text{otherwise}
\end{cases}
\quad \begin{array}{l}
m=1,\ldots,N_{\mathrm{e}},\\
n=1,\ldots,N_{\mathrm{c}}-1,
\end{array}
\end{split}
\end{equation}
where $v_m^+,\ v_m^- $ and $c_m^+,\ c_m^- $ are the pairs of vertices and cells attached to the $m^{\mathrm {th}}$ edge.
The associated quasi-Helmholtz projectors are independent of the removed column. Let \begin{equation*}
    A_f=( \Lambda_f,H_f,\Sigma_f),\quad A_g=( \Lambda_g,H_g,\Sigma_g).
\end{equation*}
Under a plane-wave excitation, the low-frequency scaling of the loop and star components of the physical traces is  \cite{Beghein2017_lf_PMCHWT}
\begin{equation}\label{physical_scaling}
\begin{split}
    A_f^{-1}(\mathbf{\hat{n}} \times \mathbf{e})^f&=(O(1),O(\omega),O(\omega))^T,\\
    A_g^{-1}(\mathbf{\hat{n}} \times \mathbf{h})^g&=(O(1),O(\omega),O(\omega))^T,
\end{split}
\end{equation}
whereas the auxiliary unknown $\mathbf p^f$ scales like a PEC current
\begin{equation}
   A_f^{-1}\mathbf p^f=(O(1),O(1),O(\omega))^T,
\end{equation}
since \eqref{singlesource} reduces to the MFIE when $\kappa_0=\kappa_1$ and $\eta_0=\eta_1$. 
If the loop and star components of the physical traces are stored together, the scaling in \eqref{physical_scaling} would cause loss of significant digits in the star components and then in the radiated far field at low frequencies \cite{Beghein2017_lf_PMCHWT}. Moreover, the physical traces computed from \eqref{physical_traces_discretized} would already be affected by significant errors if the loop and star components of the auxiliary unknown are stored together.      
\vspace{-0.3cm}
\section{Rescaling in the Primal and Dual Space}
To prevent loss of significant digits at low frequencies, the loop and star components of the auxiliary unknown and physical traces are rescaled.
Correct scaling of both quantities is essential, since loss of significant digits in the auxiliary unknown propagates to the physical traces, whereas loss of digits in the latter affects the far field computation.

The primal space quasi-Helmholtz projectors are defined by
\begin{equation}
\begin{split}
    P^\Sigma_f&=\Sigma_f (\Sigma_f^T\Sigma_f)^{-1}\Sigma_f^T,\quad P^{\Lambda H}_f=Id-P^\Sigma_f,\\
    P^{\Lambda}_f&=\Lambda_f (\Lambda_f^T\Lambda_f)^{-1}\Lambda_f^T,\quad P^H_f=P^{\Lambda H}_f-P^{\Lambda}_f,
\end{split}
\end{equation}
where $Id$ is the identity matrix. 
We also define the quasi-Helmholtz projectors in the dual space as
\begin{equation}
\begin{split}
    P^\Sigma_g=P^{\Lambda}_f,\quad P^{\Lambda H}_g=Id-P^\Lambda_f.
\end{split}
\end{equation}
These definitions reflect the duality between the RWG and BC spaces: since RWG and BC basis functions represent currents flowing across and along mesh edges, respectively, the coefficient vector associated with a local loop in the RWG space corresponds to a star in the BC space.  

The rescaling matrices in the RWG and BC spaces are
\begin{equation}
\begin{split}
    M_x = P^{\Lambda H}_x + j\frac{\kappa_0}{\kappa_*} P^\Sigma_x,\quad M_x^{-1}=P^{\Lambda H}_x-j\frac{\kappa_*}{\kappa_0} P^\Sigma_x,
\end{split}
\end{equation}
with $x \in \{f,g\}$. The parameter $\kappa_*$ is introduced for dimensional consistency. It can be chosen as $\frac{1}{d}$ where $d$ is the diameter of the geometry, and is omitted in the remainder of this work.

By inserting the rescaling matrices into \eqref{discretized_singlesource}, we obtain
\begin{equation}\label{rescaled_SSIE}
\begin{split}
&\bigg[\Big((\widetilde{\mathcal{K}}_{\kappa_0})^f_{\ f}-\frac{1}{2}Id\Big)\Big(-(\widetilde{\mathcal{K}}_{\kappa_1})^f_{\ f}+\frac{1}{2}Id\Big)+\\
&+\frac{\eta_0}{\eta_1}(\widetilde{\mathcal{T}}_{\kappa_0})^f_{\ g}(\widetilde{\mathcal{T}}_{\kappa_1})^g_{\ f}\bigg]\widetilde{\mathbf{p}}^f=M^{-1}_fN^{fg}(\mathbf{\hat{n}} \times \mathbf{e}^{\mathrm{inc}})_g,
\end{split}   
\end{equation}
where $\widetilde{\mathbf{p} }^f=M^{-1}_f\mathbf p^f$, and
\begin{equation}
    \begin{split}(\widetilde{\mathcal{K}}_{\kappa_i})^f_{\ f}&=M^{-1}_fN^{fg}(\mathcal{K}_{\kappa_i})_{gf}M_f, \\(\widetilde{\mathcal{T}}_{\kappa_i})^x_{\ y}&=M^{-1}_xN^{xy}(\mathcal{T}_{\kappa_i})_{yy}M_y.
    \end{split}
\end{equation}
The rescaled physical traces are
\begin{equation}
    \begin{split}
\label{physical_traces_rescaled}
M^{-1}_g(\mathbf{\hat{n}} \times \mathbf{h})^g&=-\frac{1}{\eta_1}(\widetilde{\mathcal{T}}_{\kappa_1})^g_{\ f}\widetilde{\mathbf{p}}^f,\\
M^{-1}_f(\mathbf{e}\times\mathbf{\hat{n}})^f&=\big(-(\widetilde{\mathcal{K}}_{\kappa_1})^f_{\ f}+\tfrac{1}{2}Id\big)\widetilde{\mathbf{p}}^f.
    \end{split}
\end{equation}
The rescaled unknown $ \widetilde{\mathbf{p}}^f$ and the physical traces scale as
\begin{equation}\label{new_scaling}
\begin{split}
    A_f^{-1}\widetilde{\mathbf{p} }^f&=(O(1),O(1),O(1))^T,\\
    A_g^{-1}M^{-1}_g(\mathbf{\hat{n}} \times \mathbf{h})^g&=(O(1),O(\omega),O(1))^T, \\
    A_f^{-1}M^{-1}_f(\mathbf{e}\times\mathbf{\hat{n}})^f&=(O(1),O(\omega),O(1))^T.
\end{split}
\end{equation}
The local-loop and star components now have equal scaling. The global-loop components of the physical traces remain $O(\omega)$, but their far-field contributions are $O(\omega)$ with respect to the contributions from the local loop and star components.
  

\begin{figure*}[!t]

\begin{subfigure}{0.33\textwidth}
\includegraphics[width=\linewidth]{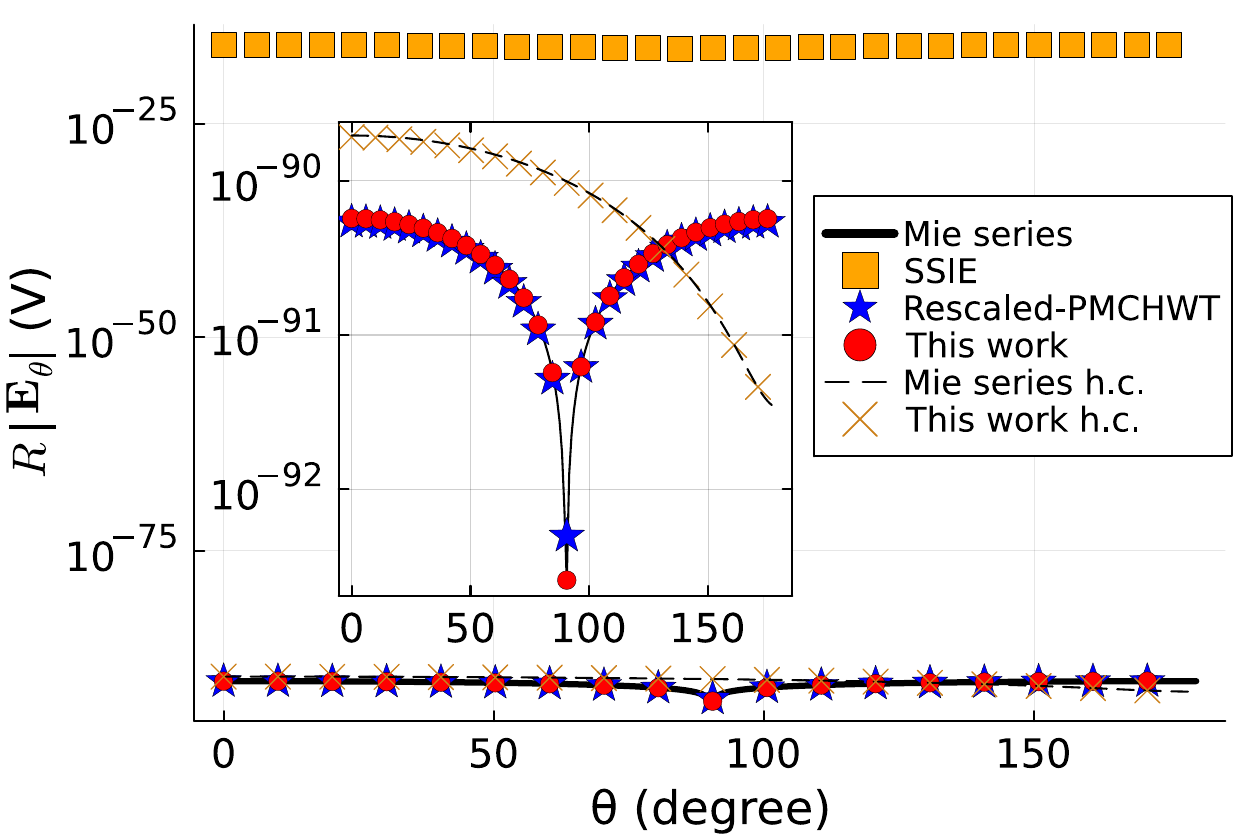} 
\caption{}
\label{fig:subim1}
\end{subfigure}
\begin{subfigure}{0.33\textwidth}
\includegraphics[width=\linewidth]{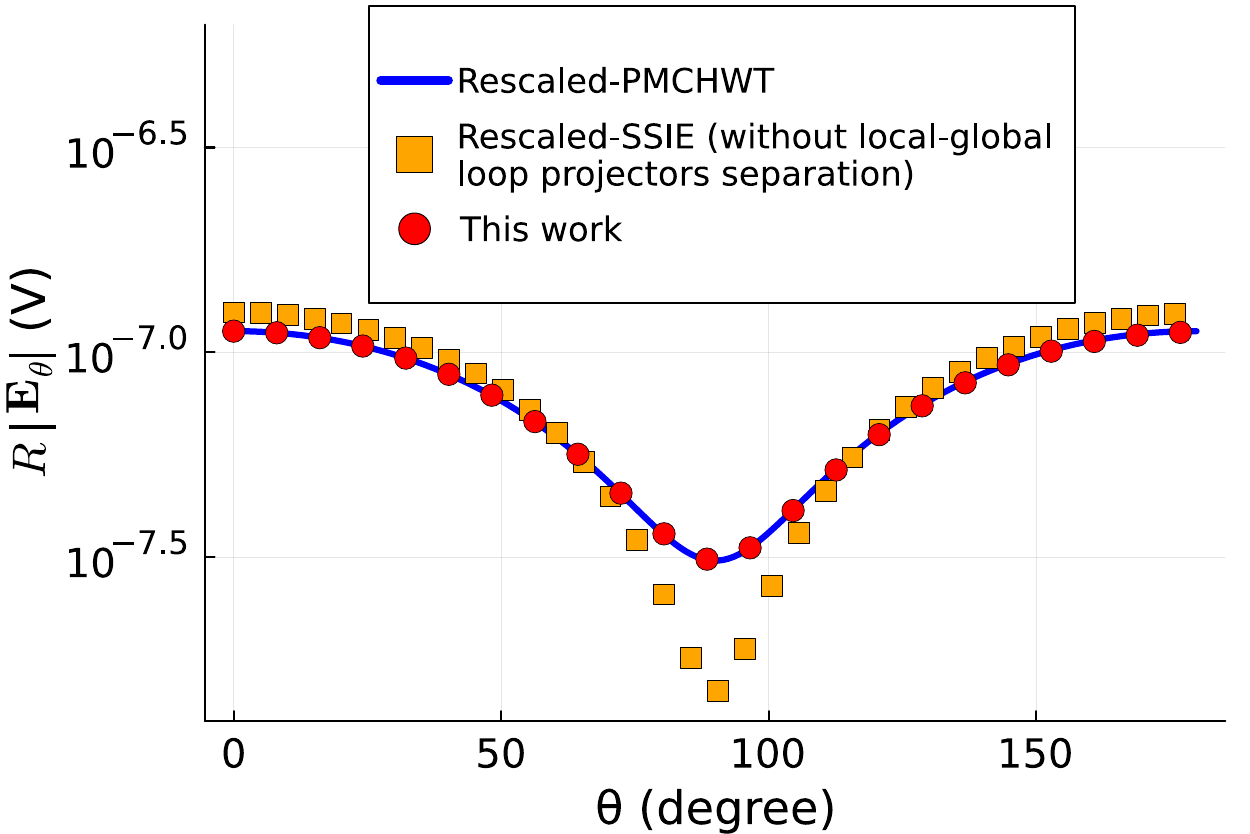}
\caption{}
\label{fig:subim2}
\end{subfigure}
\begin{subfigure}{0.33\textwidth}
  \centering
  \begin{overpic}[width=\linewidth]{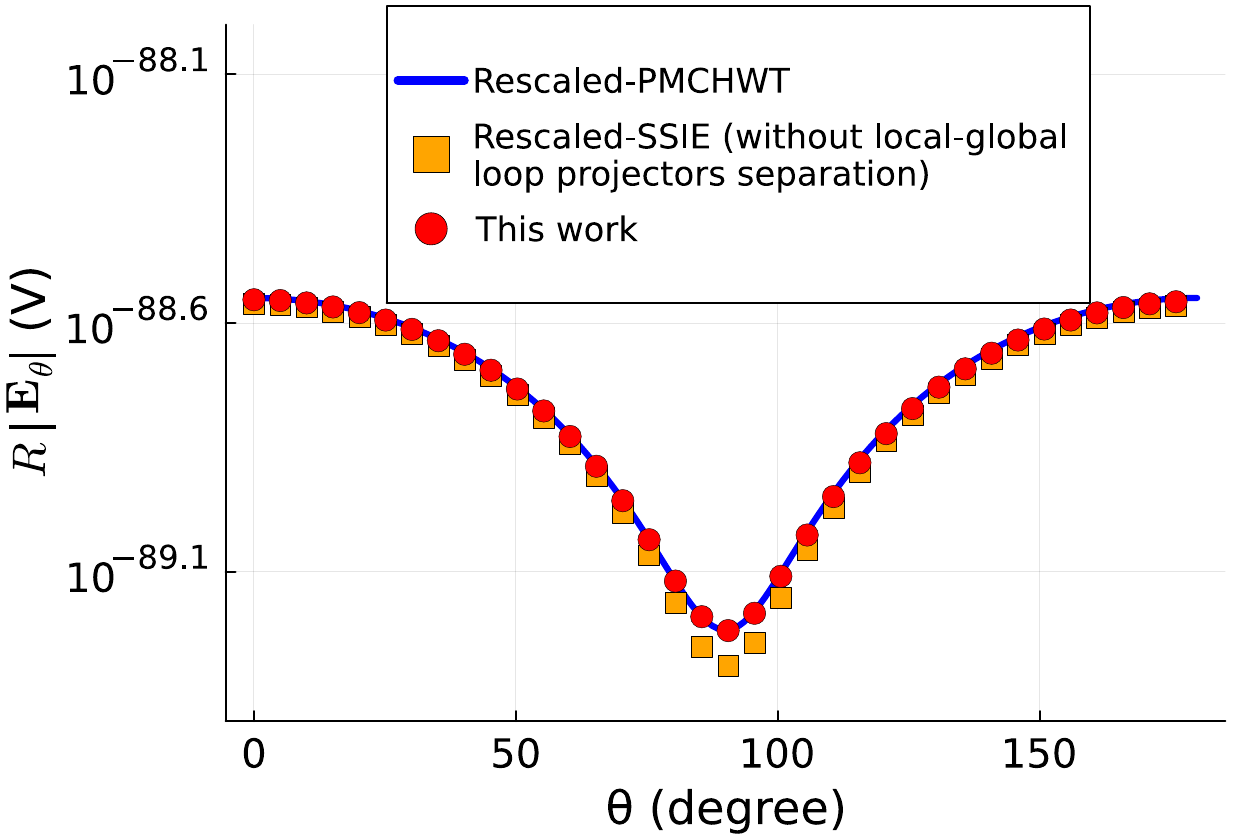}
    \put(71,10){\includegraphics[width=0.35\linewidth,trim=270 80 120 270,clip]{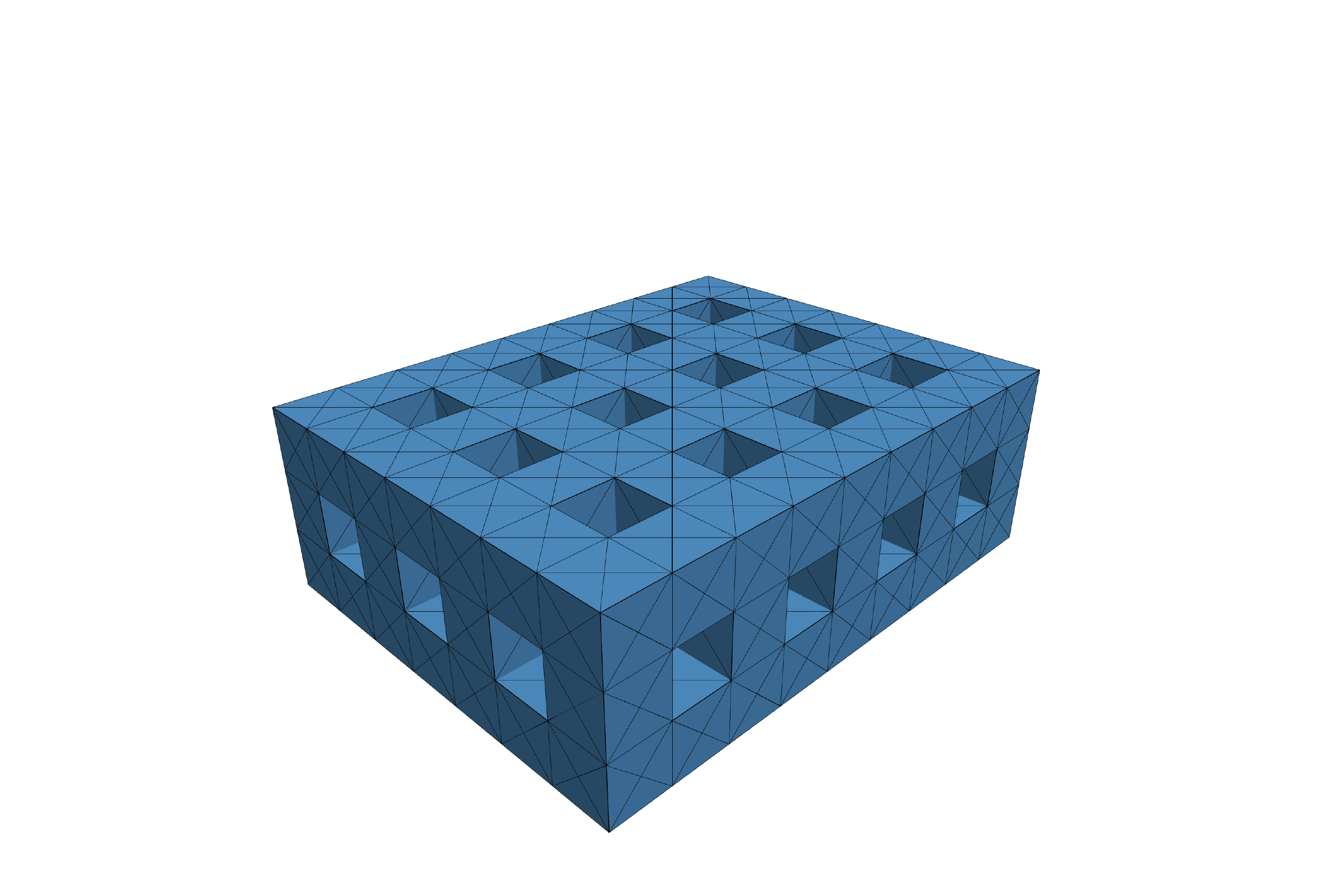}}
  \end{overpic}
  \caption{}
  \label{fig:subim3}
\end{subfigure}

\caption{Far field: magnitude $\lim_{R \rightarrow \infty} R|\mathbf{E}_\theta|$ of the $\theta$-component of the electric field at distance $R$ as a function of the angle $\theta$ in the plane containing the propagation direction and the incident electric field.
\subref{fig:subim1} Sphere: far-field results at wavenumber $\kappa_0=10^{-45}\mathrm m^{-1}$. The inset shows the curves near $10^{-91}\mathrm{V}$. Unlabeled curves refer to $\epsilon_r=5$, $\mu_r=1$; curves marked ``h.c.'' refer to the high-contrast case, $\epsilon_r=80$, $\mu_r=1000$. The regular SSIE produces a far field several orders of magnitude larger than the correct one. The rescaled SSIE agrees with the Mie series, even in the high-contrast case.
\subref{fig:subim2} Torus: far-field results at wavenumber $\kappa_0=10^{-4}\mathrm m^{-1}$. The rescaled SSIE agrees with the rescaled PMCHWT. The far-field produced when the second condition in \eqref{dbl_static_condition} is not enforced is visibly less accurate, already at moderately low frequencies. 
\subref{fig:subim3} M-cuboid g-43: far-field results at wavenumber $\kappa_0=10^{-45}\mathrm m^{-1}$. The rescaled SSIE agrees with the rescaled PMCHWT. The far field produced when the second condition in \eqref{dbl_static_condition} is not enforced is visibly less accurate.}
\label{fig:far-field results}
\vspace{-0.4cm}
\end{figure*}

Care must be taken when implementing the rescaled operators $\widetilde{\mathcal{O}}_{\kappa_i}$ and the right-hand side. 
In particular, the rescaled interior single-layer operators need to be implemented as 
\begin{align*}\label{rescaled_T_primal}
        & \!\!\!(\widetilde{\mathcal{T}}_{\kappa_1})^g_{\ f}=P^\Sigma_g N^{gf}(\mathcal{T}_{s,\kappa_1})_{ff}P^\Sigma_f-
        j\frac{1}{\kappa_0}P^\Sigma_g N^{gf}(\mathcal{T}_{s,\kappa_1})_{ff}P^{\Lambda H}_f \nonumber \\
        &+j\kappa_0 P^{\Lambda H}_g N^{gf}(\mathcal{T}_{\kappa_1})_{ff}P^\Sigma_f+P^{\Lambda H}_gN^{gf}(\mathcal{T}_{s,\kappa_1})_{ff}P^{\Lambda H}_f,\!
\end{align*}
whereas $(\widetilde{\mathcal{T}}_{\kappa_0})^f_{\ g}$ is implemented analogously by switching $f$ and $g$. This equation enforces\cite{Beghein2017_lf_PMCHWT}
\begin{equation}
(\mathcal{T}_{h,\kappa_1})_{ff}P^{\Lambda H}_f=0, \quad P^{\Sigma}_g N^{gf}(\mathcal{T}_{h,\kappa_1})_{ff}=0.
\end{equation}
The rescaled double-layer operators should be implemented as
\begin{multline*}\label{rescaled_K}
        \!\!\!\!(\widetilde{\mathcal{K}}_{\kappa_i})^f_{\ f}=P^\Sigma_f N^{fg}(\mathcal{K}_{\kappa_i})_{gf}P^\Sigma_f
        -j\frac{1}{\kappa_0}P^\Sigma_f N^{fg}(\mathcal{K}_{e,\kappa_i})_{gf}P^{\Lambda H}_f\\+j\kappa_0 P^{\Lambda H}_f N^{fg}(\mathcal{K}_{\kappa_i})_{gf}P^\Sigma_f
        +P^{\Lambda}_fN^{fg}(\mathcal{K}_{\kappa_i})_{gf}P^{\Lambda H}_f\\
        +P^{H}_fN^{fg}(\mathcal{K}_{\kappa_i})_{gf}P^{H}_f
        +P^{H}_fN^{fg}(\mathcal{K}_{e,\kappa_i})_{gf}P^{\Lambda}_f,
\end{multline*}
where $\mathcal{K}_{e,\kappa_i}=\mathcal{K}_{\kappa_i}-\mathcal{K}_{0}$ and $ \mathcal{K}_{0}$ is the static part of the double-layer operator. 
This equation enforces 
\begin{equation} \label{dbl_static_condition}
    \!\!(P^{\Sigma}_f)N^{fg}(\mathcal K_{0})_{gf}P^{\Lambda H}_f=0, \quad \!(P^{H}_f)N^{fg}(\mathcal K_{0})_{gf}P^{\Lambda}_f=0.
\end{equation} 
We recall that $\mathcal{K}_{e,\kappa_i}=O(\omega^2)$. Both conditions in \eqref{dbl_static_condition} must be enforced to ensure accurate evaluation of the $O(\omega^2)$ residual in $\mathcal{K}_{e,\kappa_i}$. The first condition has been enforced in previous formulations, e.g., the PMCHWT \cite{Beghein2017_lf_PMCHWT}, the MFIE, and the M\"uller equation \cite{LC2026}, since its violation introduces quadrature errors multiplied by ${\kappa_i}^{-1}$. In the SSIE, the second condition is also essential; otherwise, we observe significant errors in both the physical traces and the far field already at moderately low frequencies. 

The accurate evaluation of $(\mathcal{K}_{e,\kappa_i})_{gf}$ requires a numerically stable implementation to evaluate the higher-order $O(\omega^2)$ residual of the integrand.

Assuming a plane-wave excitation, the right-hand side of \eqref{rescaled_SSIE} is computed as
\begin{equation}
\begin{split}
    M^{-1}_fN^{fg}(\mathbf{\hat{n}} \times \mathbf{e}^{\mathrm{inc}})_g&=-j\frac{1}{\kappa_0}P^{\Sigma}_f N^{fg}(\mathbf{\hat{n}} \times \mathbf{e}_{\mathrm{ex}}^{\mathrm{inc}})_g+\\
    &+P^{\Lambda H}_f N^{fg}(\mathbf{\hat{n}} \times \mathbf{e}^{\mathrm{inc}})_g,
\end{split}
\end{equation}
where $\mathbf{e}_{\mathrm{ex}}^{\mathrm{inc}}=
\mathbf{\hat{e}}\,( e^{-j\kappa_0\boldsymbol{\hat{\kappa}}_0\cdot \mathbf r}-1)$. A numerically stable computation, using the $\mathrm{expm1}$ function available in e.g. Python, MATLAB, and Julia, is required.

In the static limit $\kappa_0=\kappa_1=0$, \eqref{discretized_singlesource} and \eqref{rescaled_SSIE} admit the same null space of the static MFIE on multiply connected geometries. Consequently, the operator in \eqref{rescaled_SSIE} has $N_H$ eigenvalues proportional to $O(\omega^2)$, where $N_H$ is the number of handles of the scatterer. Nevertheless, this fixed-dimensional set of eigenvalues does not affect the convergence of Krylov iterative solvers, resulting in stable iteration counts even for geometries with a large number of handles. Numerical errors might appear in the auxiliary solution $\mathbf{\widetilde{p}}^f$ due to the static null-space. However, we do not expect such errors to propagate to the physical traces, since it can be shown that the operators in \eqref{physical_traces_rescaled} admit the same static null-space as the entire left-hand side operator in \eqref{rescaled_SSIE}. Any spurious content in the auxiliary solution would be strongly suppressed in the computation of the physical traces. A detailed analysis of this numerical effect is left for future work.
\vspace{-0.25cm}
\section{Numerical Results}
We present numerical results for scattering by three geometries: a unit sphere, a torus with radii $1\mathrm m$ and $3\mathrm m$, and a topologically nontrivial level-1 Menger cuboid with dimensions $7\mathrm{m}\times9\mathrm{m}\times3\mathrm{m}$, containing $3\times4\times1$ orthogonal intersecting through-tubes, uniformly spaced, with $1\mathrm{m}\times1\mathrm{m}$ square cross sections, resulting in genus $g=43$ (M-cuboid g-43), shown in Fig.~\ref{fig:subim3}.
For the sphere, the analytical Mie series solution is used as reference. For the torus and the M-cuboid g-43, we used the rescaled PMCHWT equation presented in \cite{Beghein2017_lf_PMCHWT} as reference.
In all experiments, we set the relative permittivity and permeability $\epsilon_r=5$, $\mu_r=1$ unless otherwise specified. The incident field is a plane wave. For the torus and the M-cuboid g-43 the polarization is at $45^\circ$ with respect to the $z$-axis and the propagation direction is along the $x$-axis. The torus axis is aligned with the $z$-axis. The M-cuboid g-43 is placed with the shortest side aligned with the $z$-axis and the longest side aligned with $y$-axis.
 \begin{figure}
    \centering
        \includegraphics[width=0.9\linewidth]{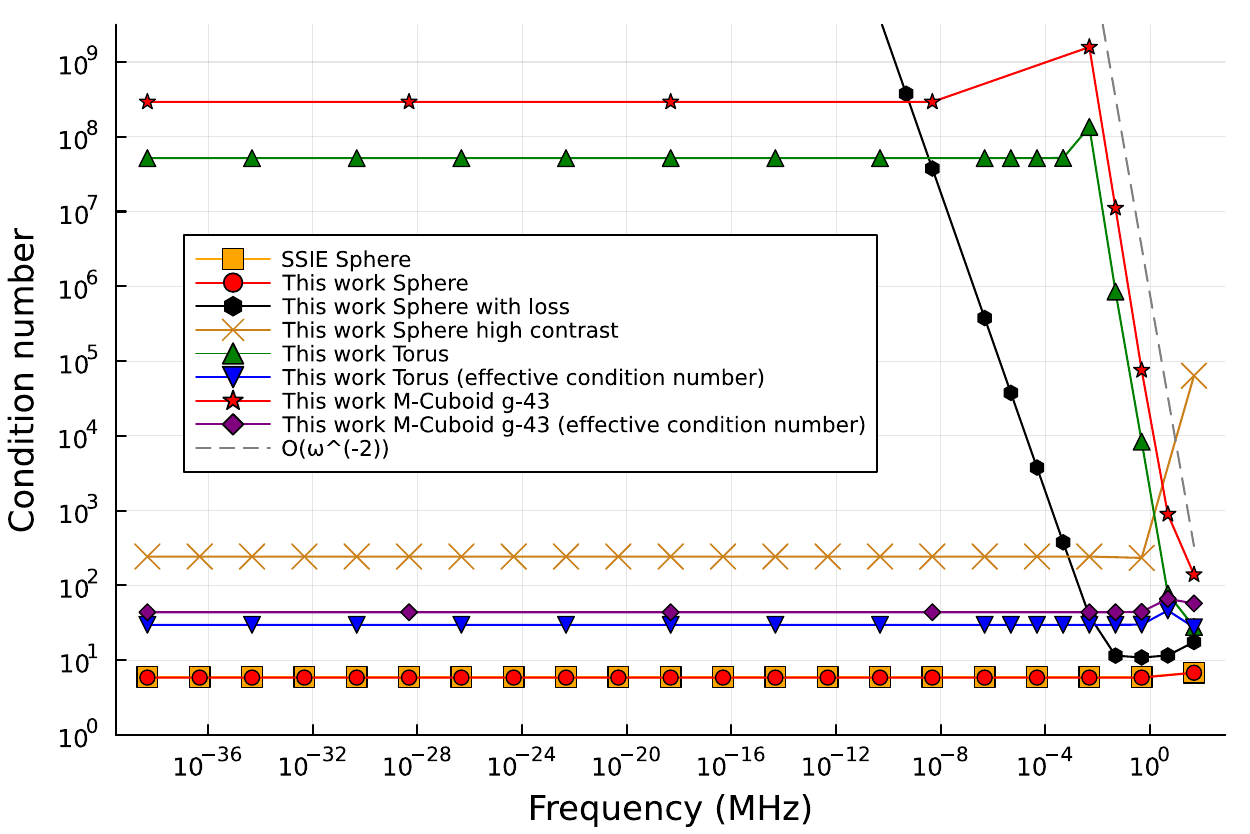}        
        \caption{
Condition number versus frequency.
The condition number is the ratio of the largest to the smallest singular value of the left-hand side operator in \eqref{rescaled_SSIE}; the effective condition number excludes the smallest $N_H$ singular values, which correspond to the static null-space. The condition number remains stable in the low-frequency regime for the lossless low-contrast and high-contrast spheres, and grows for the lossy sphere once the frequency enters the eddy-current regime. For lossless multiply connected geometries, the effective condition number remains stable.  \label{fig:condition_all}}
\vspace{-0.4cm}
\end{figure}
In Fig.~\ref{fig:far-field results} we show the far field for the three geometries. The far field computed from the solution of \eqref{physical_traces_discretized} can be several orders of magnitude different from the correct value. Instead, the far field obtained with \eqref{physical_traces_rescaled} matches the Mie series and the rescaled PMCHWT. In Figs.~\ref{fig:subim2} and \ref{fig:subim3} we also show the solution obtained when the second condition in \eqref{dbl_static_condition} is not enforced. The resulting far field is visibly less accurate than the far field obtained when both conditions in \eqref{dbl_static_condition} are enforced, already at wavenumber $\kappa_0=10^{-4}\mathrm{m}^{-1}$.

\begin{figure}
    \centering
        \includegraphics[width=0.9\linewidth]{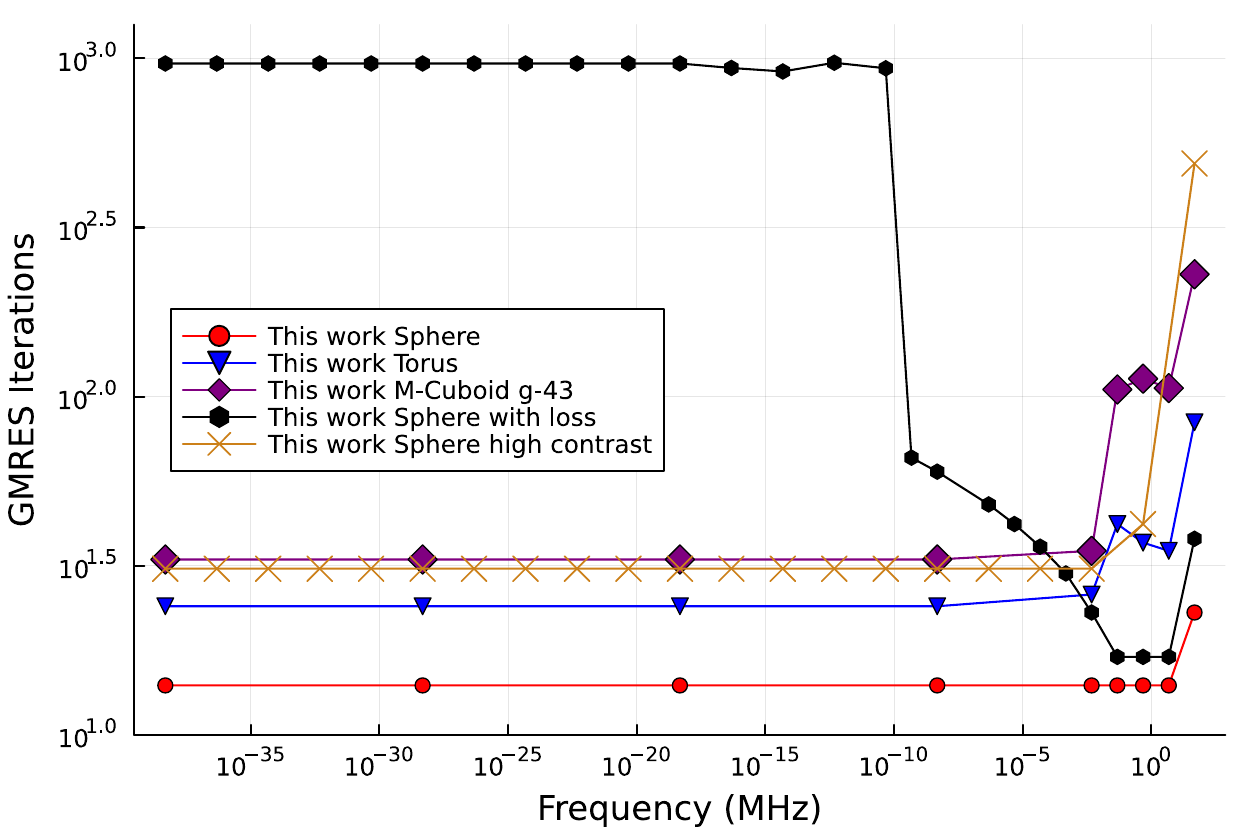}\\
        
        \caption{The number of GMRES iterations remains stable at extremely low frequencies, for all the lossless objects. For the lossy sphere, the number of iterations grows once the frequency enters the eddy-current regime.\label{fig:iters_all}}
\end{figure}

Fig.~\ref{fig:condition_all} shows the condition number of the left-hand side operator in \eqref{rescaled_SSIE} versus frequency. Fig.~\ref{fig:condition_all} also shows the results for a lossy sphere, with $\epsilon_1=\epsilon_0 ( \epsilon_r-j\frac{\sigma \eta_0}{\kappa_0})$, $\sigma=10^{-5}\,\mathrm{S/m}$, $\epsilon_r=10$, $\mu_r=1$, and a high-contrast sphere, with $\epsilon_r=80$, $\mu_r=1000$. For the lossless objects, the standard SSIE \eqref{discretized_singlesource} is already well-conditioned at all frequencies, provided that the condition $N^{fg}(\mathcal{T}_h)_{gg}N^{gf}(\mathcal{T}_h)_{ff}=0$ is enforced. The same behavior is preserved in the rescaled SSIE \eqref{rescaled_SSIE}. For multiply connected geometries with lossless materials, the condition number grows as $O(\omega^{-2})$ due to the static null-space, and stabilizes around $10^8$ because of quadrature precision.
However, the effective condition number, defined as the ratio of the largest singular value to the smallest singular value after excluding the smallest $N_H$ singular values, remains stable across all frequencies. The number of GMRES iterations follows this effective condition number (see Fig.~\ref{fig:iters_all}). 
In the lossy case, the rescaling is effective as long as $\sigma \eta_0<\kappa_0$. For $\kappa_0<\sigma \eta_0$, the internal domain enters the eddy-current regime, where the internal operators and the solution scale differently than in the lossless case \cite{GSM+2025}. This degrades the condition number and the GMRES iteration count, and a loss of digits in the solution is expected. However, in \cite{GSM+2025}, it is shown that the quasi-Helmholtz projector-based rescaling of the PMCHWT equation can be adapted to this regime. We expect a similar adaptation to be possible for the SSIE. In the high-contrast case, the rescaling remains effective: the condition number is higher than in the low-contrast case but stable in the low-frequency regime, the iteration count is mildly larger and stable down to extremely low frequencies, and the far field remains accurate (Fig.~\ref{fig:subim1}).

\section{Conclusion}
We presented a novel low-frequency stable single-source integral equation for scattering by homogeneous lossless dielectric media. By rescaling both the auxiliary unknown and the operators defining the SSIE using quasi-Helmholtz projectors, loss of significant digits is avoided at arbitrarily low frequencies while preserving the advantages of the single-source formulation. The separation between the local and global loop projectors is required, in contrast to other low-frequency stabilization techniques in the literature.
Numerical results for simply and multiply connected geometries demonstrate the accuracy, robustness, and favorable conditioning of the proposed method over a wide frequency range.
\IEEEtriggeratref{16}

\bibliographystyle{ieeetr}
\bibliography{bibsinglesource}


\end{document}